\documentclass{iopart}

\begin{document}

\jl{31}

\comment{Comment on `Dimensional expansion for the delta-function
potential'}
\author{R M Cavalcanti\footnote{E-mail: rmoritz@fma.if.usp.br}}
\address{Instituto de F\'{\i}sica, Universidade de S\~ao Paulo,
Cx.\ Postal 66318, 05315-970 S\~ao Paulo, SP, Brazil}

\begin{abstract}
I criticize the claim, made in a recent article 
(Bender C M and Mead L R 1999 {\it Eur.\ J.\ Phys.\ }{\bf 20} 117--21), 
that in order to obtain the correct 
cross section for the scattering from a 
two-dimensional delta-function potential one must perform
analytic continuation in the dimension of space.
\end{abstract}

%\submitted
%\maketitle

In a recent paper Bender and Mead \cite{Bender} obtained
the total cross section for
the scattering from an attractive delta-function
potential in $D$-dimensional space. For $D=2$, in particular,
they found
\begin{equation}
\label{sigma}
\sigma_{D=2}=\frac{4\pi^2}{k\left[\pi^2+4(\ln x)^2\right]},
\end{equation}
where $k$ is the momentum of the particle (in units 
such that $\hbar=2m=1$) and $x=(-E_0)^{1/2}/k$, where $E_0<0$ is the
energy of the (unique) bound-state. 
Bender and Mead then claimed that the results
of references \cite{Tarrach} and \cite{Mead} for the same cross section
are wrong, the cause of the error being that ``only integer values of 
$D$ were studied; analytic continuation in the
variable $D$ was not performed.''
 
The purpose of this comment is to show that (i)
the result of \cite{Tarrach} is equivalent to (\ref{sigma}),
and (ii) the result of \cite{Mead} is
wrong not because of the method employed there, but
because a limit was computed incorrectly; when properly
computed the final result is identical to (\ref{sigma}).

Let us start with \cite{Tarrach}. 
The following expression for the $s$-wave phase-shift
was obtained there:
\begin{equation}
\label{delta0}
\tan \delta_0(k)=\frac{\pi}{2\ln(k/\sqrt{-E_0})}
=-\frac{\pi}{2\ln x}.
\end{equation}
In two dimensions the total cross section can
be written in terms of the phase-shifts as \cite{Lin}
\begin{equation}
\sigma_{D=2}=\frac{4}{k}\,\sum_{m=-\infty}^{\infty}
\sin^2\delta_m(k).
\end{equation}
Using the trigonometric identity 
$\sin^2z=\tan^2z/(1+\tan^2z)$ and the fact that,
due to the zero-range character of the potential,
$\delta_m(k)=0$ for $m\ne 0$, one recovers (\ref{sigma}).

Let us now examine Mead and Godines' computation of 
$\sigma_{D=2}$. 
Inserting equations (6) and (9) in (14) (the equation
numbering here refers to \cite{Mead}) and integrating
the result over the scattering angle yields
the following expression for the total cross section:
\begin{equation}
\label{s2}
\sigma_{D=2}=\lim_{\epsilon\to 0}\frac{1}{4k}\,
\bigg|\,\frac{1}{2\pi}\,K_0(\mu\epsilon)
-\frac{i}{4}\,H_0^{(1)}(k\epsilon)\,\bigg|^{-2},
\end{equation} 
where $K_0$ is the modified Bessel function of order zero, 
$H_0^{(1)}$ is the Hankel function (of the first kind) of
order zero, and $\mu=\sqrt{-E_0}$. Neglecting terms that 
vanish when $z\to 0$, one may write \cite{GR}
\begin{equation}
\label{K0}
K_0(z)\approx -\ln\frac{z}{2}-\gamma,
\end{equation}
\begin{equation}
\label{H0}
H_0^{(1)}(z)=J_0(z)+iN_0(z)\approx 1+\frac{2i}{\pi}\left(
\ln\frac{z}{2}+\gamma\right),
\end{equation}
where $\gamma=0.57721566\ldots$ is Euler's constant.
Inserting (\ref{K0}) and (\ref{H0}) in (\ref{s2})
and taking the limit, one arrives at
\begin{equation}
\sigma_{D=2}=\frac{1}{4k}\,\bigg|\,\frac{1}{2\pi}\,
\ln\frac{k}{\mu}-\frac{i}{4}\,\bigg|^{-2}
=\frac{4\pi^2}{k\left[\pi^2+4(\ln x)^2\right]},
\end{equation}
which is precisely the result found in \cite{Bender}.
The reason why Mead and Godines did not find the correct result 
is now clear: they retained only the leading
term (i.e., $\ln z$) in the small $z$ expansion of $K_0(z)$ and 
$H_0^{(1)}(z)$, neglecting
terms which remain finite in the limit $z\to 0$.

Finally, I would like to draw attention to a recent paper
by Mitra \etal \cite{Mitra}, where the scattering from a
delta-function potential is studied using various
regularization schemes, including dimensional regularization.
The delta-function potential was also studied
in detail by Albeverio \etal \cite{Albeverio} using the method of
self-adjoint extensions. Their result for the 
$s$-wave phase-shift in two dimensions is equivalent to (\ref{delta0}).
 
\ack The author acknowledges the financial support from FAPESP.
He is also grateful to Pavel Exner and Fritz Gesztesy for calling his
attention to reference \cite{Albeverio}.


\section*{References}

\begin{thebibliography}{10}

\bibitem{Bender} Bender C M and Mead L R 1999 Dimensional expansion
for the delta-function potential {\it Eur. J. Phys.} {\bf 20} 117--21

\bibitem{Tarrach} Gosdzinsky P and Tarrach R 1991 Learning quantum
field theory from elementary quantum mechanics {\it Am. J. Phys.}
{\bf 59} 70--4

\bibitem{Mead} Mead L R and Godines J 1991 An analytical example of
renormalization in two-dimensional quantum mechanics {\it Am. J. Phys.}
{\bf 59} 935--7

\bibitem{Lin} Lin Q-G 1997 Levinson theorem in two dimensions
{\it Phys. Rev.} A {\bf 56} 1938--44

\bibitem{GR} Gradhsteyn I S and Ryzhik I M 1965 {\it Tables of
Integrals, Series, and Products} (New York: Academic)

\bibitem{Mitra} Mitra I, DasGupta A and Dutta-Roy B 1998
Regularization and renormalization in scattering from Dirac
delta potentials {\it Am. J. Phys.} {\bf 66} 1101--9

\bibitem{Albeverio} Albeverio S, Gesztesy F, H{\o}egh-Krohn R
and Holden H 1988 {\it Solvable Models in Quantum Mechanics}
(New York: Springer-Verlag)

\end{thebibliography}
\end{document}